\documentclass[11pt]{amsart}
\usepackage{amsmath}
\usepackage{epsfig}
\usepackage{verbatim}
\usepackage{amssymb}
\newcommand{\eproof}{\hfill\rule{2.2mm}{3.0mm}}

\newcommand{\Proof}{\noindent {\bf Proof.~~}}

\newcommand{\R}{{\mathbb R}}

\newcommand{\C}{{\mathbb C}}

\newcommand{\Prob}{{\mathbb P\,}}
\newcommand{\Normal}{{\mathcal N}(0,1)}
\newcommand{\Span}{{\rm span}}
\newcommand{\ep}{\varepsilon}

\renewcommand{\eqref}[1]{(\ref{#1})}
\newcommand{\inner}[1]{\langle #1 \rangle}
\newcommand{\shsp}{\hspace{1em}}
\newcommand{\mhsp}{\hspace{2em}}

\newcommand{\va}{{\mathbf a}}

\newcommand{\vv}{{\mathbf v}}
\newcommand{\vu}{{\mathbf u}}
\newcommand{\vf}{{\mathbf f}}

\newcommand{\HH}{{\mathbf H}}

\newcommand{\F}{{\mathcal F}}

\newtheorem{prop}{Proposition}[section]
\newtheorem{lem}[prop]{Lemma}

\newtheorem{theo}[prop]{Theorem}

\begin{document}
\baselineskip 18pt
\title{Random Matrices and Erasure Robust Frames}
\author{Yang Wang}
\thanks{Yang Wang was supported in part by the
       National Science Foundation grant DMS-08135022 and DMS-1043032.}
\address{Department of Mathematics  \\ Michigan State University\\
East Lansing, MI 48824, USA}
\email{ywang@math.msu.edu}
\begin{abstract}
    Data erasure can often occur in communication. Guarding against erasures involves
    redundancy in data representation. Mathematically this may be achieved by redundancy
    through the use of frames. One way to measure the robustness of a frame against erasures
    is to examine the worst case condition number of the frame with a certain number of
    vectors erased from the frame. The term {\em numerically erasure-robust frames (NERFs)} was
    introduced in \cite{FicMix12} to give a more precise characterization of erasure robustness
    of frames. In the paper the authors established that random frames whose entries are drawn
    independently from the standard normal distribution can be robust against up to approximately
    15\% erasures, and asked whether there exist frames that are robust against erasures of
    more than 50\%. In this paper we show that with very high probability
    random frames are, independent of the dimension, robust against any amount of erasures
    as long as the number of remaining vectors is at least $1+\delta$ times the dimension for some
    $\delta_0>0$. This is the best possible result, and it also implies that the proportion of
    erasures can arbitrarily close to 1 while still maintaining robustness. Our result depends
    crucially on a new estimate for the smallest singular value of a rectangular random matrix
    with independent standard normal entries.
\end{abstract}
\maketitle

\section{Introduction}
\setcounter{equation}{0}

Let $\HH$ be a Hilbert space. A set of elements $\F=\{\vf_n\}$ in $\HH$
(counting multiplicity) is called a {\em frame} if there exist two positive constants
$C_*$ and $C^*$ such that for any $\vv\in {\HH}$ we have
\begin{equation}  \label{1.1}
     C_* \|\vv\|^2 \leq \sum_n |\inner{\vv,\vf_n}|^2 \leq C^* \|\vv\|^2.
\end{equation}
The constants $C_*$ and $C^*$ are called the {\em lower frame bound} and
the {\em upper frame bound}, respectively. A frame is called a {\em tight
frame} if $C_*=C^*$. In this paper we focus mostly on real finite
dimensional Hilbert spaces with $\HH =\R^n$ and $\F=\{\vf_n\}_{j=1}^N$,
although we shall also discuss the extendability of the results to the
complex case.
Let $F=[\vf_1,\vf_2, \dots,\vf_N]$. It is
called the {\em frame matrix} for $\F$.
It is well known that $\F$ is a frame if and only if the $n\times N$ matrix $F$ has rank $n$.
Furthermore, the optimal frames bounds are given by
$$
     C_* = \sigma_n^2(F), \mhsp C^* = \sigma_1^2(F),
$$
where $\sigma_1 \geq \sigma_2 \geq \cdots \sigma_n>0$ are the singular
values of $F$. Throughout this paper we shall identify
without loss of generality a frame $\F$ by its
frame matrix.

The main focus of the paper is on the {\em erasure robustness} property for a frame.
This property arise in applications such as communication where data can
be lost or corrupted in the process of transmission. Suppose that we have a frame $F$ that
is {\em full spark} in the sense that every $n$ columns of $F$ span $\R^n$,
it is theoretically possible to erase up to $N-n$ data from the full set of data
$\{\inner{\vv,\vf_j}\}_{j=1}^N$ while still reconstruct the signal $\vv$.
This is a simple consequence of the property that with the remaining
available data $\{\inner{\vv,\vf_j}\}_{j\in S}$ with $|S|\geq n$,
$\vv$ is uniquely determined because $\Span (\vf_j: j\in S) =\R^n$.
In practice, however, the condition number of the matrix $[\vf_j]_{j\in S}$
could be so poor that the reconstruction is numerically unstable against
the presence of additive noise in the data. Thus robustness against data loss and erasures
is a highly desirable property for a frame. There have been a number of studies
 that aim to address this important issue.

Among the first studies of erasure-robust frames was given in \cite{GoKoKe01}.
It was shown in subsequent
studies that that unit norm tight frames are optimally robust against one erasure \cite{CasKov03}
while Grassmannian frames are optimally robust against two
erasures \cite{StrHea03,HolPau04}. The literature on erasure robustness for frames is
quite extensive, see e.g. also \cite{KoDrGo02,Vers05,PusKov05}.
In general, the robustness of a frame $F$ against $q$-erasures, where
$q\leq N-n$, is measured by the maximum of the condition numbers of all $n\times (N-q)$
submatrices of $F$. More precisely, let $S \subseteq \{1,2, \dots, N\}$ and let
$F_S$ denote the
$n\times |S|$ submatrix of $F$ with columns $\vf_j$ for $j\in S$ (in its natural order, although the
order of the columns is irrelevant). Then the robustness against $q$-erasures of $F$ is measured by
\begin{equation} \label{1.1a}
      R(F,q) := \max_{|S|=N-q} \frac{\sigma_1(F_S)}{\sigma_n(F_S)}.
\end{equation}
Of course, the smaller $R(F,q)$ is the more robust $F$ is against $q$-erasures.
In \cite{FicMix12}, Fickus and Mixon coined the term {\em numerically erasure robust frame (NERF)}. A frame
$F$ is $(K,\alpha,\beta)$-NERF if
$$
     \alpha\leq \sigma_n(F_S)\leq \sigma_1(F_S)\leq \beta \mhsp \mbox{for any}~~S\subseteq \{1,2,\dots, N\},~|S|=K.
$$
Thus in this case $R(F,N-K) \leq \beta/\alpha$. Note that for any full spark $n\times N$
frame matrix $F$ and any $n\leq K\leq N$ there always exist $\alpha,\beta>0$ such
that $F$ is $(K,\alpha,\beta)$-NERF. The main goal is to find classes of frames
where the bounds $\alpha, \beta$, and more importantly, $R(F,N-K)=\beta/\alpha$,
are independent of the dimension $n$ while allowing the proportion of erasures
$1-\frac{K}{N}$ as large as possible. The authors studied in \cite{FicMix12}
the erasure robustness of
$F=\frac{1}{\sqrt n} A$, where the entries of $A$ are independent random variables
of the standard normal $\Normal$ distribution.
It was shown that with high probability
such a matrix can be good NERFs provided that $K$ is no less than approximately
$85\%$ of $N$. The authors also proved that equiangular
frame $F$ in $\C^n$ with $N=n^2-n+1$ vectors is a good NERF against up to about $50$\% erasures.
As far as the proportion of erasures
is concerned this was the best known result for NERFs. However, the frame requires
almost $n^2$ vectors. The authors posed as an open question whether
there exist NERFs with $K < N/2$. A more recent paper \cite{FJMP12}
explored a deterministic construction based on certain group theoretic techniques.
The approach offers more flexibility in the frame design than the far more restrictive
equiangular frames.

In this paper we revisit the robustness of random frames. We provide a much stronger result
for random frames, showing
that for any $\delta>0$, with very high probability, the frame $F=\frac{1}{\sqrt n} A$
is a $((1+\delta)n, \alpha,\beta)$-NERF where $\alpha,\beta$
depend only on $\delta$ and the aspect ratio $\frac{N}{n}$.
One version of our result is given by the following theorem.

\begin{theo} \label{theo-1.1}
    Let $F=\frac{1}{\sqrt n} A$ where $A$ is $n\times N$ whose entries are independent
Gaussian random variables of $N(0,1)$ distribution. Let $\lambda = \frac{N}{n}>1$. Then for
any $0<\delta_0<\lambda-1$ and $\tau_0>0$ there exist $\alpha, \beta>0$ depending
only on $\delta_0, \lambda$ and $\tau_0$ such that for any
$\delta_0 \leq \delta<\lambda-1$, the frame $F$ is a $((1+\delta)n, \alpha, \beta)$-NERF
with probability at least $1-e^{-\tau_0 n}$.
\end{theo}

Later in the paper we shall provide more implicit estimates for $\alpha,\beta$ that
will allow us to easily compute them numerically. Note that our result
is essentially the best possible, as we cannot go to $\delta_0=0$. A corollary
of the theorem is that for random Gaussian frames the proportion of erasures
$1-\frac{K}{N}$ can be made arbitrary large
while the frames still maintain robustness with overwhelming probability.

Our theorem depends crucially on a refined estimate on the smallest singular value of
a random Gaussian matrix. There is a wealth of literature on random matrices. The study
of singular values of random matrices has been particularly intense
in recent years due to their applications in compressive sensing for the construction of
matrices with the so-called {\em restricted isometry property}
(see e.g.\cite{CanTao05,CanTao06,BDDW08,Cand08}). Random matrices have also been
employed for phase retrieval \cite{CESV11}, which aims to reconstruct a signal from
the magnitudes of its samples.
For a very informative and
comprehensive survey of the subject we refer the readers to \cite{RudVer10,Vers10},
which also contains an extensive list of references (among the notable ones
\cite{Edel88,RudVer09,TaoVu10}). For the $n\times N$ Gaussian random matrix $A$ the expected value of
$\sigma_1(A)$ and $\sigma_n(A)$ are asymptotically $\sqrt N +\sqrt n$ and
$\sqrt N -\sqrt n$, respectively. Many important
results, such as the NERF analysis of random matrices in \cite{FicMix12} as well as results
on the restricted isometry
property in compressive sensing, often utilize known estimates of
$\sigma_1(A)$ and $\sigma_n(A)$ based on Hoeffding-type
inequalities. One good such estimate is
\begin{equation} \label{1.2}
      \Prob (\sigma_n(A) < \sqrt N -\sqrt n -t) \leq e^{-\frac{t^2}{2}},
\end{equation}
see \cite{Vers10}. The problem with this estimate is that even by taking $t=\sqrt N -\sqrt n$
we only get a bound of $e^{-(\sqrt\lambda-1)^2n/2}$ even though the probability in this case
is 0. Thus estimates such as (\ref{1.2}) that cap the decay rate are often inadequate.
When applied to the erasure robustness
problem for frames they usually put a cap on the proportion of erasures.
To go further we must prove an estimate that will allow the exponent
of decay to be much larger. We achieve this goal by proving the following theorem:

\begin{theo} \label{theo-1.2}
    Let $A$ be $n\times N$ whose entries are independent
random variables of standard normal $N(0,1)$ distribution.
Let $\lambda = \frac{N}{n}>1$. Then for any $\mu>0$
there exist constants $c, C>0$ depending only on $\mu$ and $\lambda$ such that
\begin{equation} \label{1.3}
    \Prob\bigl(c\sqrt n\leq\sigma_n(A)\leq \sigma_1(A)\leq C\sqrt n\bigr) \geq 1- 3e^{-\mu n}.
\end{equation}
Furthermore, we may take $C=1+\sqrt\lambda+\sqrt\mu$ and $c = \sup_{0<t<1} \varphi(t)$ where
\begin{equation} \label{1.4}
    \varphi(t) =\frac{t^{\frac{1}{\lambda}}}{L}  -\frac{2Ct}{1-t},
        \mhsp\mbox{where}\mhsp L=\sqrt{\frac{2e}{\lambda}}e^{\frac{\mu}{\lambda}}.
\end{equation}
\end{theo}

\noindent
{\bf Acknowledgement.}~The author would like to thank Radu Balan and Dustin Mixon for very helpful discussions.

\section{Smallest Singular Value of a Random Matrix: Nonasymptotic Estimate}
\setcounter{equation}{0}

We begin with estimates on the extremal singular values of a ranodm matrix $A$ whose entries are
independent standard normal random variables. We shall assume throughout the section that
$A$ is $n\times N$ where $\frac{N}{n} = \lambda >1$. One of the very important estimates is
\begin{equation} \label{2.1}
   \Prob \left(\sigma_1(A) > \sqrt N +\sqrt n +t\right) \leq e^{-\frac{t^2}{2}},
\end{equation}
see \cite{Vers10}. Our main goal of this section is to prove the estimates for smallest
singular value $\sigma_n(A)$
stated in Theorem \ref{theo-1.2}. An equivalent formulation of (\ref{2.1}) is
\begin{equation} \label{2.2}
   \Prob \bigl(\sigma_1(A) > C\sqrt n\bigr)
      \leq e^{-\frac{(C-1-\sqrt\lambda)^2}{2}n},\mhsp C \geq 1+\sqrt\lambda.
\end{equation}
Observe that
$$
     \sigma_n(A) = \min_{\vv\in S^{n-1}} \|A^*\vv\|,
$$
where $S^{n-1}$ denotes the unit sphere in $\R^n$.

\begin{lem}  \label{lem-2.1}
   Let $c>0$. For any $\vv\in S^{n-1}$ the probability $\Prob(\|A^*\vv\|\leq c)$ is
   independent of the choice of $\vv$. We have
\begin{equation} \label{2.3}
   \Prob \bigl(\|A^*\vv\| \leq \sqrt{\delta n}\bigr) \leq \Bigl(\frac{2e\delta}{\lambda}\Bigr)^{\frac{N}{2}}
\end{equation}
for any $\delta>0$.
\end{lem}
\Proof The fact that $\Prob(\|A^*\vv\|\leq c)$ is independent of the
choice of $\vv$ is a well know fact, which stems from the fact that the entries of $PA$ are again
independent standard normal random variables for any orthogonal $n\times n$ matrix $P$. In
particular, one can always find an orthogonal $P$ such that $P\vv=e_1$. Thus we may
without loss of generality take $\vv=e_1$. In this case
$\|A^*\vv\|^2 = a_{11}^2 + \cdots +a_{1N}^2$ where $[a_{11}, \dots, a_{1N}]$
denotes the first row of $A$.
Denote $Y_N =  a_{11}^2 + \cdots +a_{1N}^2$. Then $Y_N$ has the $\Gamma(\frac{N}{2}, 1)$ distribution, which has the density function
$$
     \rho (t) = \frac{1}{\Gamma(\frac{N}{2})}e^{-t} t^{\frac{N}{2}-1}, \shsp t>0.
$$
Denote $m=\frac{N}{2}$. It follows that
\begin{eqnarray*}
    \Prob \bigl(\|A^*\vv\| \leq \sqrt{\delta n}\bigr)
          &=&  \Prob \bigl(Y_N \leq \delta n\bigr)  \\
          &=& \frac{1}{\Gamma(m)}\int_{0}^{\delta n} e^{-t} t^{m-1} dt \\
          &\leq& \frac{1}{\Gamma(m)}\int_{0}^{\delta n} t^{m-1} dt \\
          &=& \frac{\delta^m n^m}{\Gamma(m)}.
\end{eqnarray*}
Note that $\Gamma(m) \geq (\frac{m}{e})^m$ by Stirling's formula. The theorem now follows from
$\frac{N}{n}=\lambda$ and $m = \frac{N}{2}$.
\eproof

A ubiquitous tool in the study of random matrices is an $\ep$-net. F
or any $\ep>0$ an $\ep$-net for $S^{n-1}$ is a set in $S^{n-1}$ such that any point on
$S^{n-1}$ is no more than $\ep$ distance away from the set. The following result
is known and can be found in \cite{Vers10}:

\begin{lem}  \label{lem-2.2}
    For any $\ep>0$ there exists an $\ep$-net ${\mathcal N}_\ep$ in $S^{n-1}$ with cardinality no larger
than $(1+2\ep^{-1})^n$.
\end{lem}

\vspace{2mm}

\noindent
{\bf Proof of Theorem~\ref{theo-1.2}.}~~Assume that $\sigma_n(A) = b\sqrt n$. Then there exists a
$\vv_0 \in S^{n-1}$ such that $\|A^*\vv_0\| = b\sqrt n$. Let ${\mathcal N}_\ep$ be an $\ep$-net
for $S^{n-1}$ and take $\vu \in {\mathcal N}_\ep$ that is the closest to $\vv_0$. So
$\|\vu -\vv_0\| \leq \ep$. Thus
\begin{equation} \label{2.4}
     \|A^*\vu\| \leq \|A^*\vv_0\| + \|A^*(\vu-\vv_0)\| \leq b\sqrt n + \ep \sigma_1(A).
\end{equation}
Hence
\begin{equation} \label{2.5}
     \Prob\bigl(\sigma_n(A)\leq c\sqrt n\bigr) \leq \sum_{\vu\in {\mathcal N}_\ep}
                    \Prob\bigl(\|A^*\vu\|\leq c\sqrt n+\ep\sigma_1(A)\bigr).
\end{equation}
Note that
\begin{eqnarray*}
    \Prob\bigl(\|A^*\vu\|\leq c\sqrt n+\ep\sigma_1(A)\bigr)
    &=&\Prob\bigl(\|A^*\vu\|\leq c\sqrt n+\ep\sigma_1(A), \sigma_1(A) \leq C\sqrt n\bigr)\\
    &+&\Prob\bigl(\|A^*\vu\|\leq c\sqrt n+\ep\sigma_1(A), \sigma_1(A) > C\sqrt n\bigr).
\end{eqnarray*}
By Lemma \ref{lem-2.1} the first term on the right hand side is bounded from above by
\begin{align*}
    \Prob\bigl(\|A^*\vu\|&\leq c\sqrt n+\ep\sigma_1(A), \sigma_1(A) \leq C\sqrt n\bigr)\\
       &\leq \Prob\bigl(\|A^*\vu\|\leq c\sqrt n+\ep C\sqrt n\bigr)
                  \leq \Bigl(\frac{2e(c+\ep C)^2}{\lambda}\Bigr)^{\frac{N}{2}}.
\end{align*}
By (\ref{2.2}) the second term on the right hand side is bounded from above by
\begin{align*}
   \Prob\bigl(\|A^*\vu\|&\leq c\sqrt n+\ep\sigma_1(A), \sigma_1(A) > C\sqrt n\bigr) \\
       &\leq \Prob\bigl(\sigma_1(A) > C\sqrt n\bigr) \leq e^{-\frac{(C-1-\sqrt\lambda)^2}{2}n}.
\end{align*}
Thus combining these two upper bounds we obtain the estimate
\begin{equation} \label{2.6}
     \Prob\bigl(\sigma_n(A)\leq c\sqrt n\bigr) \leq
     \Bigl(1+\frac{2}{\ep}\Bigr)^n
     \left(\Bigl(\frac{2e(c+\ep C)^2}{\lambda}\Bigr)^{\frac{N}{2}} +
      e^{-\frac{(C-1-\sqrt\lambda)^2}{2}n}\right).
\end{equation}

We would like to bound $\Prob\bigl(\sigma_n(A)\leq c\sqrt n\bigr)$ by $2e^{-\mu n}$.
All we need then is to choose
$\ep, c, C>0$ so that both upper bound terms in (\ref{2.6}) are bounded by $e^{-\mu n}$.
Note that $\frac{N}{2} = \frac{\lambda}{2}n$. Hence we only need
\begin{eqnarray}
    -\mu &\geq& \ln(1+2\ep^{-1}) +
         \frac{\lambda}{2}\Bigl(\ln2e -\ln\lambda+2\ln(c+\ep C)\Bigr), \label{2.7} \\
     -\mu &\geq& -\frac{1}{2}(C-1-\sqrt \lambda)^2. \label{2.8}
\end{eqnarray}
The equation (\ref{2.8}) leads to the condition
\begin{equation}  \label{2.9}
    C \geq \sqrt {2\mu} +\sqrt\lambda +1.
\end{equation}
To meet condition (\ref{2.7}) we set $c=r\ep$. Then $\ln(c+\ep C)=-\ln\ep^{-1}+\ln(r+C)$. Thus
(\ref{2.7}) becomes
\begin{equation}  \label{2.10}
    (\lambda-1)\ln(\ep^{-1})
       \geq \mu +\ln(2+\ep)+\frac{\lambda}{2}\ln\Bigl(\frac{2e(r+C)^2}{\lambda}\Bigr).
\end{equation}
Clearly, once we fix $C$ and $r$, say, take $C=\sqrt {2\mu} +\sqrt\lambda +1$ and $r=1$,
$\ln\ep^{-1}$ will be greater than the right hand side of (\ref{2.10}) for small enough
$\ep$ because of the condition $\lambda>1$. Both $C,c$ only depend on $\lambda$
and $\mu$. The existence part of the theorem is thus proved.

While we have already a good explicit estimate $C=\sqrt {2\mu} +\sqrt\lambda +1$,
it remains to establish the explicit formula for $c$. For any fixed $r$ the largest $\ep$
is achieved when (\ref{2.10}) is an equality, namely
$$
    (\lambda-1)\ln(\ep^{-1})
       = \mu +\ln(2+\ep)+\frac{\lambda}{2}\ln\Bigl(\frac{2e(r+C)^2}{\lambda}\Bigr),
$$
which one can rewrite as
$$
    \ln(r+C) = -(1-p)\ln\ep -p \ln(2+\ep)-\ln L,
$$
where $p=\lambda^{-1}$ and $L=\sqrt{\frac{2e}{\lambda}}e^{\frac{\mu}{\lambda}}$.
It follows that
$$
   r\ep = \frac{1}{L}\Bigl(\frac{\ep}{2+\ep}\Bigr)^p -C\ep =\frac{1}{L} t^{\frac{1}{\lambda}} -\frac{2Ct}{1-t},
$$
where $t = \frac{\ep}{2+\ep}$. Note that $0<t<1$. Now we can take $c$ to be the supreme value of $r\ep$, which yields
\begin{equation}  \label{2.11}
     c = \sup_{0<t<1} ~\Bigl\{\frac{t^{\frac{1}{\lambda}}}{L} -\frac{2Ct}{1-t}\Bigr\}.
\end{equation}

Finally, (\ref{1.3}) follows from $\Prob\bigl(\sigma_n(A)\leq c\sqrt n\bigr)
\leq 2e^{-\mu n}$ and (\ref{2.2}).
The proof of the theorem is now complete.
\eproof

\vspace{2mm}
\noindent
{\bf Remark.}~~Although there does not seem to exist an explicit formula for $c$
given in (\ref{2.11}), there is a very good explicit approximation of it. In general,
the $t$ that maximize $\varphi(t)$ is rather small. So we may approximate
$\frac{2Ct}{1-t}$ simply by $2Ct$ and find the maximum of
\begin{equation}  \label{2.12}
     \tilde\varphi(t) = \frac{1}{L} t^{\frac{1}{\lambda}} -2Ct.
\end{equation}
The maximum of $\tilde\varphi(t)$ is obtained at
$t_0=(2C\lambda L)^{-\frac{\lambda}{\lambda-1}}$. This $t_0$ is very close
to the actual $t$ that maximizes $\varphi(t)$. Thus
\begin{equation}  \label{2.13}
     \tilde c:=\varphi(t_0) = \Bigl(\frac{1}{2C\lambda L^\lambda}\Bigr)^{\frac{1}{\lambda-1}}
           \Bigl(1-\frac{1}{\lambda}\Bigr)
\end{equation}
has $\tilde c \leq c$ and it is a close approximation of the optimal $c$. Of course,
Theorem \ref{theo-1.2} still holds when $c$ is replaced by $\tilde c$.

\vspace{2mm}
Although Theorem~\ref{theo-1.2} is for real Gaussian random matrices, a complex version of
it can also be proved with minor modifications. A complex random variable $Z=X+iY$ has the
{\em complex standard normal} distribution if both $X$ and $Y$ have the real complex normal
distribution $\Normal$. Theorem~\ref{theo-1.2} extends to the following theorem for
the complex case:

\smallskip

\begin{theo} \label{theo-2.3}
    Let $A$ be $n\times N$ whose entries are independent
random variables of complex standard normal $N(0,1)$ distribution.
Let $\lambda = \frac{N}{n}>1$. Then for any $\mu>0$
there exist constants $c, C>0$ depending only on $\mu$ and $\lambda$ such that
\begin{equation} \label{2.14}
    \Prob\bigl(c\sqrt n\leq\sigma_n(A)\leq \sigma_1(A)\leq C\sqrt n\bigr) \geq 1- 3e^{-\mu n}.
\end{equation}
Furthermore, we may take $C=\sqrt2+2\sqrt\lambda+2\sqrt{\mu}$ and $c = \sup_{0<t<1} \varphi(t)$ where
\begin{equation} \label{2.15}
    \varphi(t) =\frac{t^{\frac{1}{\lambda}}}{L}  -\frac{2Ct}{1-t},
        \mhsp\mbox{where}\mhsp L=\sqrt{\frac{2e}{\lambda}}e^{\frac{\mu}{2\lambda}}.
\end{equation}
\end{theo}
\Proof~The proof follows the same argument as in the real case so we only sketch the proof here.
In particular we point out the places where the estimates need to be modified.

Write $A=A_R+iA_I$ and set $B=[A_R, A_I]$. Then $B$ is an $n\times 2N$ matrix whose entries are
independent real standard normal random variables. It is easy to check that
$\sigma_1(A) \leq \sqrt2\sigma_1(B)$. Thus by taking $C=2\sqrt\lambda +\sqrt 2+2\sqrt{\mu}$
we have via (\ref{2.1}) that
\begin{equation} \label{2.16}
    \Prob\bigl(\sigma_1(A)\leq C\sqrt n\bigr) \leq e^{-\mu n}.
\end{equation}

The estimate for $\sigma_n(A)$ follows from the same strategy as in the real case.
First of all, just like the real case for any $n\times n$ unitary matrix $U$ the entries of
$UA$ are still independed complex standard normal random variables. As a result the
probability $\Prob (\|A^*\vv|| \leq \sqrt{\delta n}$) where $\vv\in\C^n$ is a unit vector
does not depend on the choice of $\vv$. By taking $\vv=e_1$ we see that $(\|A^*\vv||^2$ has
the $\Gamma(N,1)$ distribution (as opposed to the
$\Gamma(\frac{N}{2},1)$ distribution in the real case). Applying Lemma \ref{lem-2.1} we
obtain the equivalent result for the complex case in
\begin{equation} \label{2.17}
   \Prob \bigl(\|A^*\vv\| \leq \sqrt{\delta n}\bigr) \leq \Bigl(\frac{2e\delta}{\lambda}\Bigr)^N.
\end{equation}
Next for the $\ep$-net, we observe that the unit sphere in $\C^n$ is precisely the
unit sphere in $\R^{2n}$ if we identify $\C^n$ as $\R^{2n}$. Thus we can find an $\ep$-net
${\mathcal N}_\ep$ of cardinality no more than $(1+2\ep^{-1})^{2n}$. The proof of
Theorem \ref{theo-1.2} now goes through with some minor modifications. The most important one
is that with (\ref{2.16}) and (\ref{2.17}) the inequality condition  (\ref{2.7}) now becomes
$$
   -\frac{\mu}{2} \geq \ln(1+2\ep^{-1}) +
         \frac{\lambda}{2}\Bigl(\ln2e -\ln\lambda+2\ln(c+\ep C)\Bigr),
$$
where the constant $C$ is changed to $C=2\sqrt\lambda +\sqrt 2+2\sqrt{\mu}$. Substituting
this $C$ and $\frac{\mu}{2}$ for $\mu$ we prove the theorem.
\eproof

\section{Random Frames as NERFs}
\setcounter{equation}{0}

Our goal in this section is to establish the robustness of random frames against erasures
by proving Theorem~\ref{theo-1.1}. Here we restate Theorem \ref{theo-1.1} in a a different
form for the benefit of simpler notation in the proof.

\begin{theo}  \label{theo-3.1}
     Let $F=\frac{1}{\sqrt n} A$ where $A$ is $n\times N$ whose entries are drawn
     independently from the standard normal $\Normal$ distribution. Let
     $\lambda = \frac{N}{n}>1$ and $K=pN=p\lambda n$ where $\lambda^{-1}< p \leq 1$.
     For any $\tau_0>0$ there exist constants $\alpha, \beta>0$ depending only on
     $\lambda$, $p$ and $\tau_0$ such that $F$ is a $(K, \alpha, \beta)$-NERF with
     probability at least $1-3e^{-\tau_0 n}$.
\end{theo}
\Proof There exists exactly $\frac{N!}{K!(N-K)!}$ subsets $S\subseteq \{1,2,\dots,N\}$
of cardinality $|S|=K$. It is well known that
$$
    \frac{N!}{K!(N-K)!} \leq \frac{N^N}{K^K(N-K)^{N-K}},
$$
which can be shown easily by Stirling's Formula or induction on $N$.
Set $s_p = p\ln p^{-1} +(1-p)\ln (1-p)^{-1}$, which has $0\leq s_p \leq \ln2$. We have then
\begin{equation}  \label{3.1}
    \frac{N!}{K!(N-K)!} \leq \left(p^{-p}(1-p)^{p-1}\right)^N = e^{\lambda s_p n}.
\end{equation}

Now we set $\mu := \lambda s_p +\tau_0$. Let $C=\sqrt{ 2\mu} +\sqrt{p\lambda} +1$ and
$c = \sup_{0<t<1} \varphi(t)$ where $\varphi(t)$ is given in (\ref{1.4}).
Let the columns of $A$ be $\{\va_j\}_{j=1}^N$.
For any $S\subseteq \{1,2,\dots,N\}$ we denote by $A_S$ the
submatrix of $A$ whose columns are $\{\vf_j:~j\in S\}$. Then for
$|S|=K=p\lambda n$ we have
$$
    \Prob\left(c\sqrt n\leq\sigma_n(A_S)\leq\sigma_1(A_S)\leq C\sqrt n\right)
    \geq 1- 3e^{-\mu n}.
$$
by Theorem \ref{theo-1.2}. It follows that
\begin{align*}
    \Prob\Bigl(\sigma_n(A_S) \leq c\sqrt n~~&\mbox{or}~~\sigma_1(A_S)\geq C\sqrt n~~\mbox{for some $S$ with $|S|=K$}\Bigr) \\
    & \leq \sum_{|S| = K} \Prob\Bigl(\sigma_n(A_S) \leq c\sqrt n~~\mbox{or}~~\sigma_1(A_S)\geq C\sqrt n\Bigr) \\
    & \leq 3 e^{(\lambda s_p -\mu) n}~=~3e^{-\tau_0 n}.
\end{align*}
It follows that
$$
    \Prob\Bigl(c\sqrt n \leq \sigma_n(A_S) \leq \sigma_1(A_S)\leq C\sqrt n
     ~~\mbox{for all $S$ with $|S|=K$}\Bigr)
    \geq 1-3e^{\tau_0 n}.
$$
This implies that, by setting $\alpha=c$ and $\beta=C$, $F=\frac{1}{\sqrt n} A$ is a $(K,\alpha,\beta)$-NERF
with probability at least $1-3e^{-\tau_0 n}$.
\eproof

\vspace{2mm}

Theorems \ref{1.1} and \ref{theo-3.1} states that random Gaussian frames can be
robust with overwhelming
probability against erasures of an arbitrary proportion of data from the original data,
at least in theory, as long as the number of remaining vectors is at least $(1+\delta_0)n$
for some $\delta_0>0$. In practice one may ask how good the condition numbers are if the erasures reach a high proportion, say, 90\% of the data. We show some numerical results
below.

\vspace{2mm}

\begin{figure}
\begin{center}
\includegraphics[width=7.2cm,height=6cm]{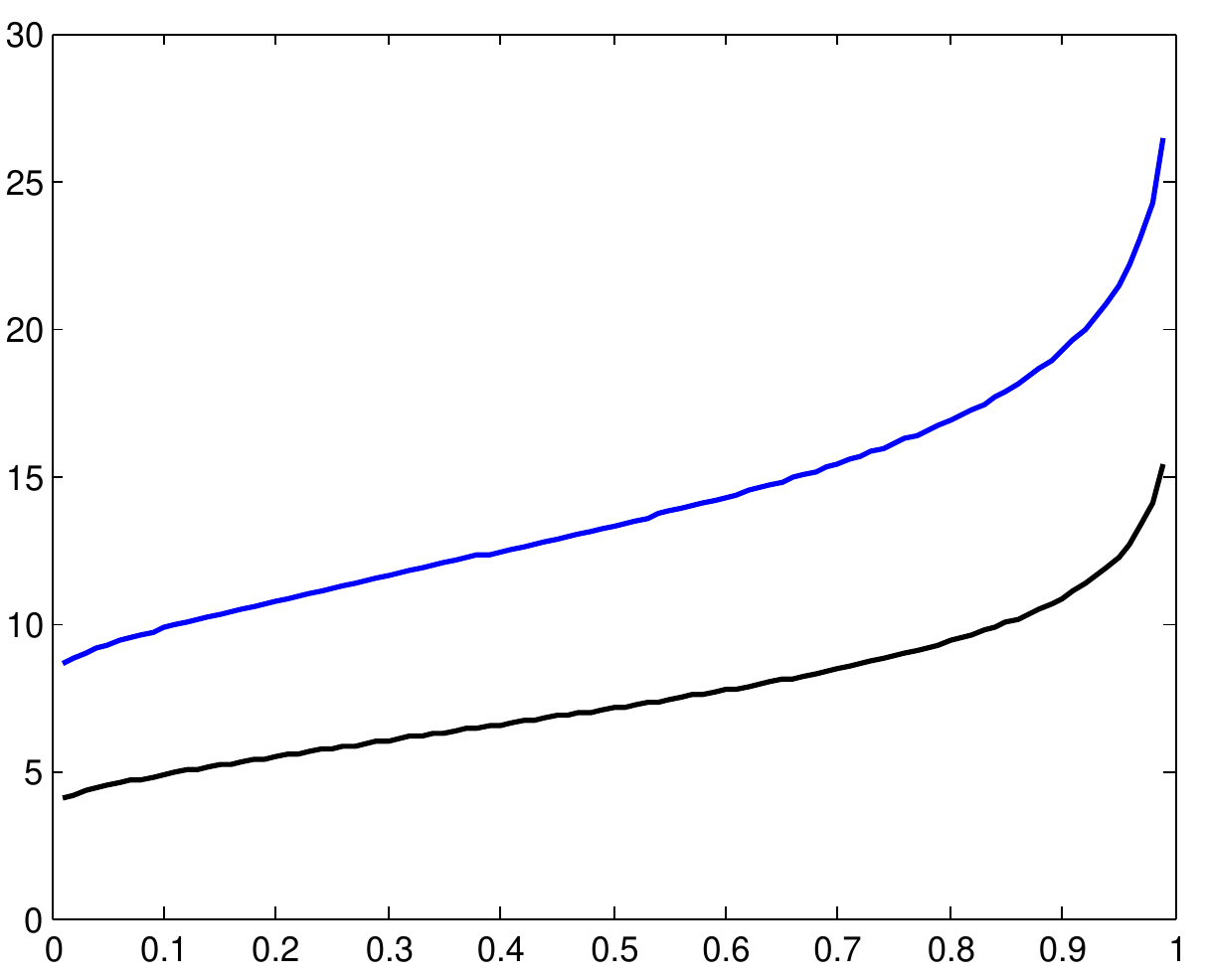}~~
\includegraphics[width=7.2cm,height=6cm]{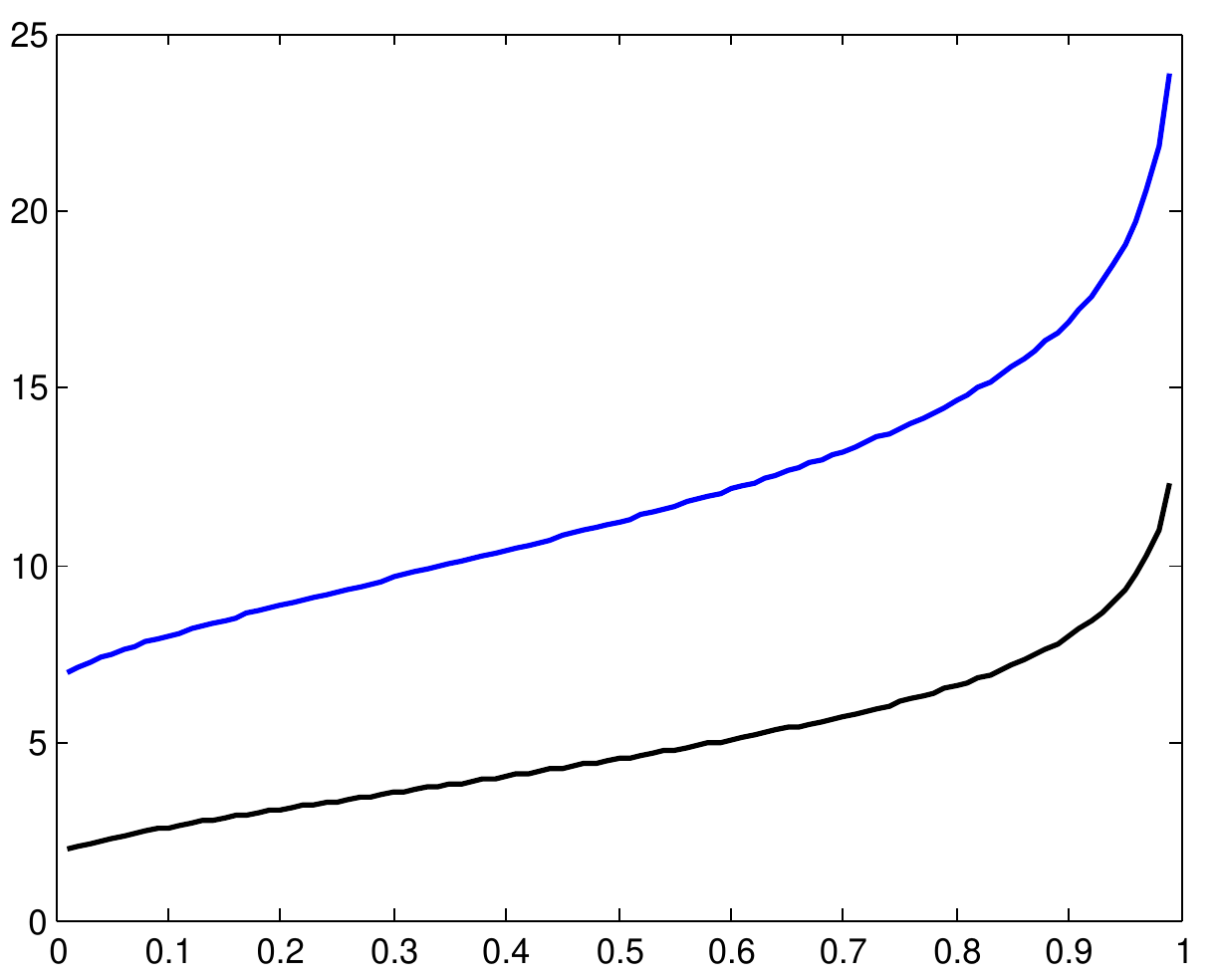}
\end{center}
\caption{\label{fig-1}Left: $\log_2(\beta/\alpha)$ against the proportion of erasures
when $N$ varies from $K$ to $100K$ while $K$ is fixed at $K=2n$
(top curve) and $K=5n$ (bottom curve).
Right: Same as in the left figure, but for $-\log_2(\alpha)$.}
\end{figure}

\noindent
{\bf Example 1.}~~Let $F=\frac{1}{\sqrt{n}}
A$ where $A$ is $n\times N$ whose
entries are independent standard normal random variables. Set $\tau_0=0.25$.
In this experiment we fix $K=2n$ and $K=5n$,
respectively, and let $N$ vary. As $N$ increases from $N=K$ to $N=100K$ the
proportion of erasure $s = 1 -\frac{K}{N}$ increases from 0 to 99\%.
We shall use $\beta/\alpha$ as a measure of robustness
since it is an upper bound for the condition number. Clearly,
as $s$ increases we should expect $\beta/\alpha$ to increase. The left plot in Figure~\ref{fig-1} shows
$\log_2(\beta/\alpha)$ against $s$ for both $K=2n$ (top curve) and $K=5n$ (bottom curve).
Because the frame is normalized so that each column is on average a unit norm vector, it
also makes sense to use the smallest singular value as a measurement of robustness.
The right plot in Figure~\ref{fig-1} shows
$-\log_2(\alpha)$ against $s$ also for both $K=2n$ (top curve) and $K=5n$ (bottom curve).
Our numerical results show that in the case $K=2n$, with probability at least $1-3e^{-0.5n}$,
the condition number is no more than $10232$ for $50$\% erasures and no more than
$611675$ for 90\% erasures. In the case $K=5n$, the corresponding numbers
are 139.88 and 1862.1, respectively. In fact, even with 99\% erasures the condition number
is no more than 42716.

\vspace{2mm}

\begin{figure}
\begin{center}
\includegraphics[width=7.2cm,height=6cm]{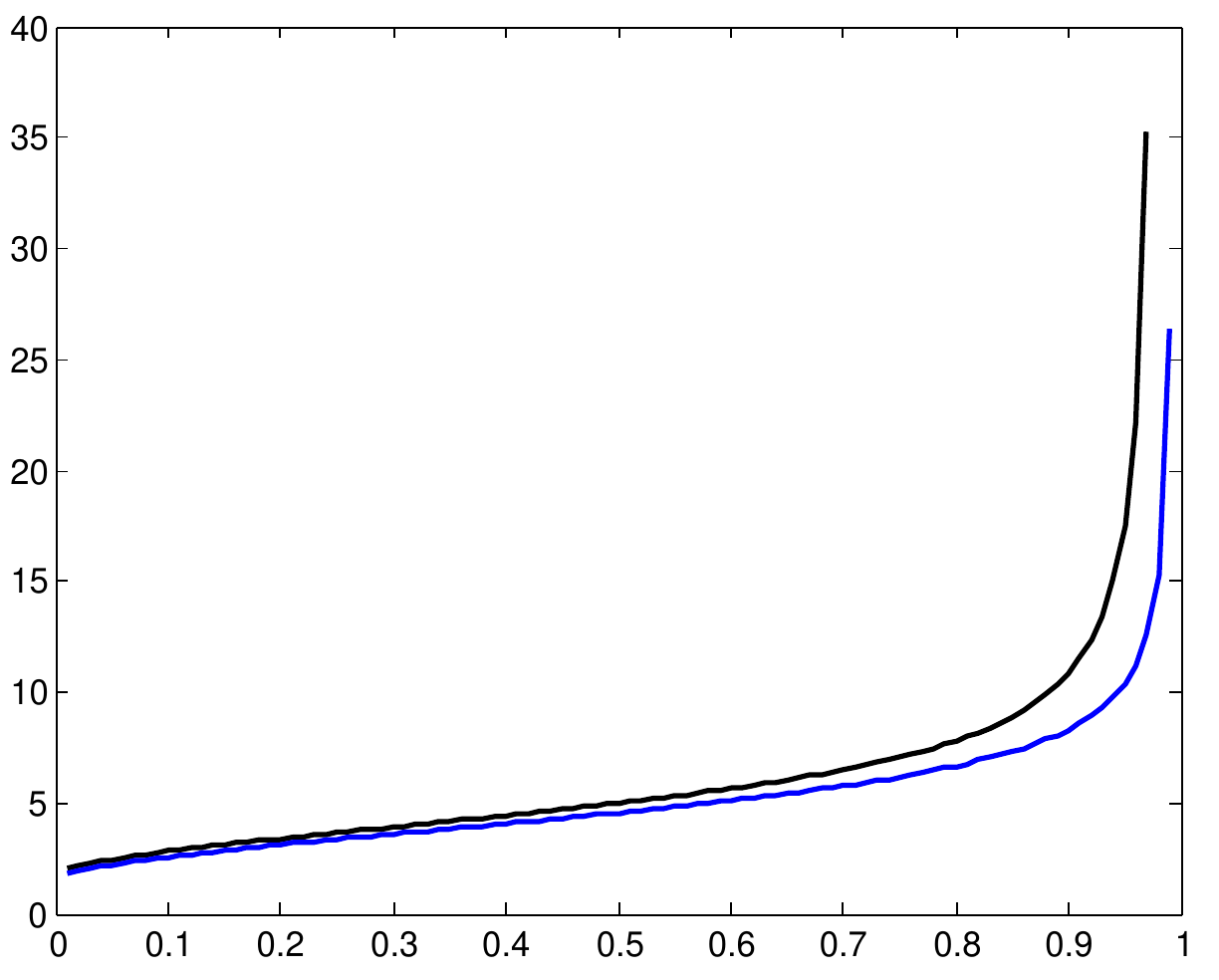}~~
\includegraphics[width=7.2cm,height=6cm]{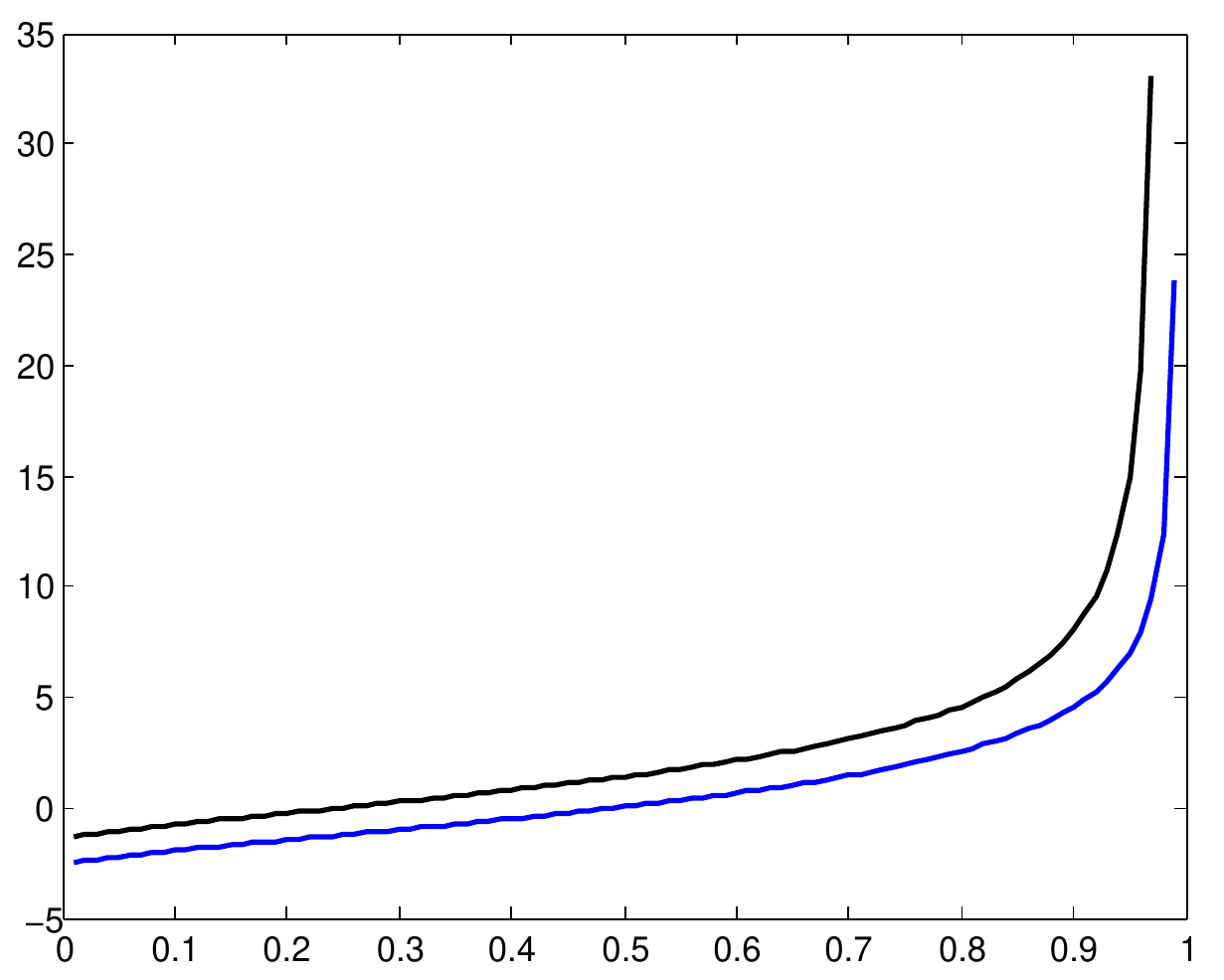}
\end{center} \caption{\label{fig-2}Left: $\log_2(\beta/\alpha)$ against the
proportion of erasures
when $K$ varies while $N$ is fixed at $N=50n$ (top curve) and $N=200n$ (bottom curve).
Right: Same as in the left figure, but for $-\log_2(\alpha)$.}
\end{figure}

\noindent
{\bf Example 2.}~~Again we let $F=\frac{1}{\sqrt{n}}A$ where $A$ is $n\times N$ whose
entries are independent standard normal random variables, and let $\tau_0=0.2
5$.
In this experiment we fix $N=200n$ and $N=50n$,
respectively, and let $K$ vary so the proportion of erasures $s=1-\frac{K}{N}$
varies from 0 to 99\% ($N=200n$ and 0 to 97\% ($N=50n$), respectively.
Again we should expect the robustness to go down as we increase $s$.
The left plot in Figure~\ref{fig-2} shows
$\log_2(\beta/\alpha)$ against $s$ for $N=50n$ (top curve) and $N=200n$ (bottom curve).
The right plot in Figure~\ref{fig-2} shows
$-\log_2(\alpha)$ against $s$ also for both $N=50n$ (top curve) and $N=200n$ (bottom curve).
Our numerical results show that in the case $N=50n$, with probability at least $1-3e^{-0.5n}$,
the condition number is no more than $31.7$ for $50$\% erasures and
$1862.1$ for 90\% erasures. In the case $N=200n$, the corresponding numbers
are 23.48 and 315.12, respectively. Even with 95\% erasures the condition number is
no more than 1312.4.

\end{document}